\documentclass{article}
\usepackage{spconf,amsmath,graphicx,bm,url,}

\title{Audio-Visual Speech Enhancement and Separation by utilizing Multi-Modal Self-Supervised Embeddings}
\begin{document}
%
\maketitle
\begin{abstract}
AV-HuBERT, a multi-modal self-supervised learning model, has been shown to be effective for categorical problems such as automatic speech recognition and lip-reading. This suggests that useful audio-visual speech representations can be obtained via utilizing multi-modal self-supervised embeddings. Nevertheless, it is unclear if such representations can be generalized to solve real-world multi-modal AV regression tasks, such as audio-visual speech enhancement (AVSE) and audio-visual speech 
separation (AVSS). In this study, we leveraged the pre-trained AV-HuBERT model followed by an SE module for AVSE and AVSS. Comparative experimental results demonstrate that our proposed model performs better than the state-of-the-art AVSE and traditional 
audio-only SE models. In summary, our results confirm the effectiveness of our proposed model for the AVSS task with proper fine-tuning strategies, demonstrating that multi-modal self-supervised embeddings obtained from AV-HuBERT can be generalized to audio-visual regression tasks.
\end{abstract}
\begin{keywords}
Audio-Visual Speech Enhancement, Audio-Visual Speech Separation, AV-HuBERT
\end{keywords}
\section{Introduction}
\label{sec:intro}

Speech enhancement (SE) and speech separation (SS) aim to extract speech signals of interest from a given utterance mixed with unwanted audio signals. With recent developments in deep learning (DL), DL-based methods have demonstrated better results than traditional SE and SS methods, either for audio-only or audio-visual (AV) applications ~\cite{Xu2015,lu13_interspeech,Hou2018,Chuang2022,gao2021VisualVoice,lee2021looking}. 
Nevertheless, most DL-based AVSE and AVSS models have their own specific modules designed to better integrate the audio-visual information for the target task, which may not be favorable from the viewpoints of some current DL model design philosophies. One popular learning paradigm is designing a unified scheme that can learn generalizable representations with minor model modifications for different tasks. Self-supervised learning (SSL) is a popular learning strategy for this purpose. SSL uses the data itself for its own learning supervision. It has been effective in several multi-modal applications, such as vision-and-language ~\cite{NEURIPS2019_c74d97b0} and audio-visual learning~\cite{shi2022avhubert,Chan2022}. Specifically, for audio-visual applications~\cite{shi2022avhubert,Chan2022}, a transformer model~\cite{NIPS2017_3f5ee243} is pre-trained with audio-visual data and then fine-tuned for classification tasks, such as automatic speech recognition (ASR) and lip-reading.  \\
\indent In this study, we investigated whether pre-trained audio-visual models are beneficial in regression-based speech processing tasks, namely AVSE and AVSS. In particular, we consider AV-HuBERT~\cite{shi2022avhubert} our audio-visual SSL model and combine AV-HuBERT with a simple neural regression model for AVSE and AVSS tasks. The results show that model performances can be improved, indicating that pre-trained audio-visual representations from AV-HuBERT are beneficial in audio-visual regression tasks.  \\
\indent The paper is organized as follows. Section~\ref{sec:related_work} gives some review on related works. Section~\ref{sec:method} presents the details of the proposed approach, and Section~\ref{sec:exp} demonstrates our experiments and  results. A summary of the paper is presented in Section~\ref{sec:conclusion}.

\section{Related research}
\label{sec:related_work}

Recently, an increasing number of audio-only self-supervised models have been applied to SE and SS. For applications in SE, some studies ~\cite{Huang2022,tsai-etal-2022-superb,Sivaraman2022} utilized pretrained latent representations as the SE model input and boosted the performance by incorporating cross-domain features~\cite{2204.03339}. For applications in SS, the authors of ~\cite{2010.15366} proposed a self-supervised pre-training approach to stabilize label assignments when training SS models. \\
\indent There are relatively few studies on self-supervised pre-training for audio-visual regression-based tasks ~\cite{shi2022avhubert,Chan2022}. AV-HuBERT~\cite{shi2022avhubert} is an extension of the Hubert~\cite{Hsu2021} model (an audio-only SSL model) for multimodal pre-training. The authors in ~\cite{Shi2022-gu} leveraged pre-learned representations from a pretrained AV-HuBERT for speaker classification and verification tasks. In contrast, we focused on two regression-based speech processing tasks, namely AVSE and AVSS.

\section{Proposed method}
\label{sec:method}

\begin{figure} [t]
\centerline{\includegraphics[scale=.33]{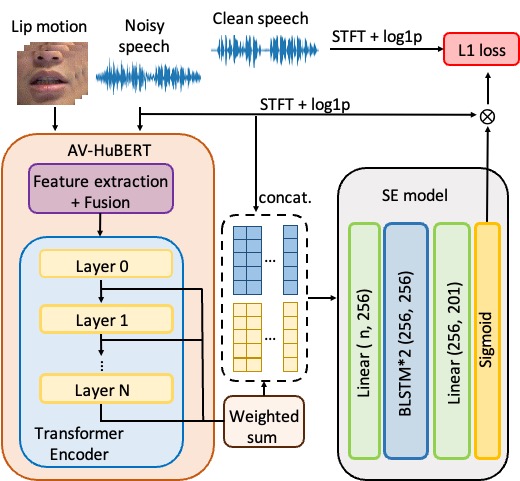}}
\caption{Our proposed AVSE model is based on the AV-HuBERT, followed by a SE model.}
\label{avse_model}
\end{figure}

\subsection{Audio-Visual SE Model}
The proposed AVSE approach leverages the AV-HuBERT model as an upstream processor, as shown in Fig.~\ref{avse_model}. The lip image sequence $V_{1:T}$ and noisy speech $A^n_{1:T}$ are fed into the AV-HuBERT; the representations from each layer of the transformer encoder are denoted as $H^l$, where $0 \leq l \leq N$, and $N$ is the number of layers. Inspired by ~\cite{Huang2022, tsai-etal-2022-superb}, a trainable function $w(\cdot)$ is applied to the representations from all layers as follows: 

\begin{equation}\label{E:weightedsum}
{H}_{\text{WS}} = \sum_{l=0}^{L}w(l)\, {H}^{(l)}
\end{equation}
where $w(l)$ denotes the weight of the $l$-th layer and has the properties $w(l) \geq 0 $ and $ \sum_{l} w(l) = 1$. ${H}_{\text{WS}}$ are then concatenated with the $log1p$ spectrogram feature from the noisy speech. The concatenated features are subsequently fed into a neural SE module consisting of fully connected layers (FC) and a two-layer bidirectional long short-term memory module (BLSTM). The output of the SE module is a soft mask and is multiplied by the magnitude of the spectrogram of the noisy speech. The training objective is to minimize the $L1$ distance between the multiplied spectrogram and that generated from the clean speech.

\subsection{Audio-Visual SS Model}
\begin{figure}[t]
\centerline{\includegraphics[scale=.30]{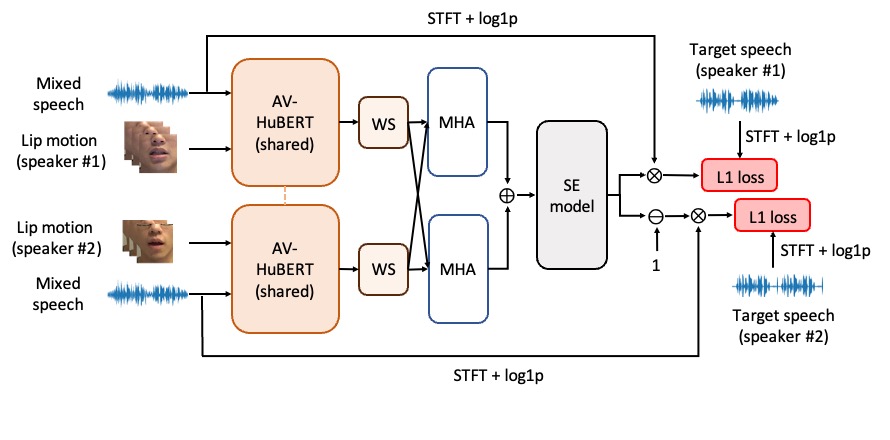}}
\caption{Proposed audio-visual SS framework. The output features of the shared AV-HuBERT from the two speakers are coupled via cross-attention and then are fed into the neural regression model for SS. The WS and MHA blocks represent the weighted-sum and the multi-head attention modules, respectively.}
\label{avsp_model}
\end{figure}

Similar to the proposed AVSE model, our proposed AVSS model uses the AV-HuBERT model as the front-end module to process audio-visual inputs. Fig.~\ref{avsp_model} shows the overall architecture of the AVSS system. The two image sequences for the speaker lip movements in a two-speaker video are denoted as $V^{s1}_{1:T}$ and $V^{s2}_{1:T}$. The mixed speech is expressed as $A^m_{1:T}$. A shared AV-Hubert model is used to generate the multi-modal representations for each speaker, followed by a weighted-sum operation, which is the same as Eq.\ref{E:weightedsum}. The resulting outputs are denoted as ${H}_{\text{WS,sp1}}$ and $H_{\text{WS,sp2}}$. To better couple the multi-modal features obtained from the two speakers, we applied cross-attention over ${H}_{\text{WS,sp1}}$ and $H_{\text{WS,sp2}}$ with a layer of multi-head attention (MHA) mechanism, which is expressed as:
\begin{equation}
\begin{split}
O_{sp1} = MHA(H_{\text{WS,sp1}},H_{\text{WS,sp2}})\\
O_{sp2} = MHA(H_{\text{WS,sp2}},H_{\text{WS,sp1}})
\end{split}
\end{equation}
The outputs of the MHA modules are then summed and fed into the SE module, which has the same architecture as the previous AVSE model. The objective of the loss is to minimize the $L1$ distances between the masked spectrograms and target spectrograms, which can be expressed as:

\begin{equation}
\begin{split}
L_{avss} = dist(SE(O_{sp1} \bigoplus O_{sp2})\bigotimes Sm, Ssp1) \\
 + dist(( 1-SE(O_{sp1} \bigoplus O_{sp2}))\bigotimes Sm, Ssp2) 
\end{split}
\end{equation}
where $SE$, $Sm$, and $S_{sp1}$ and $S_{sp2}$ denote the SE model, magnitude of the spectrograms of the mixed speech, and speech from each speaker, respectively.

\begin{figure}[t]
\centerline{\includegraphics[scale=.32]{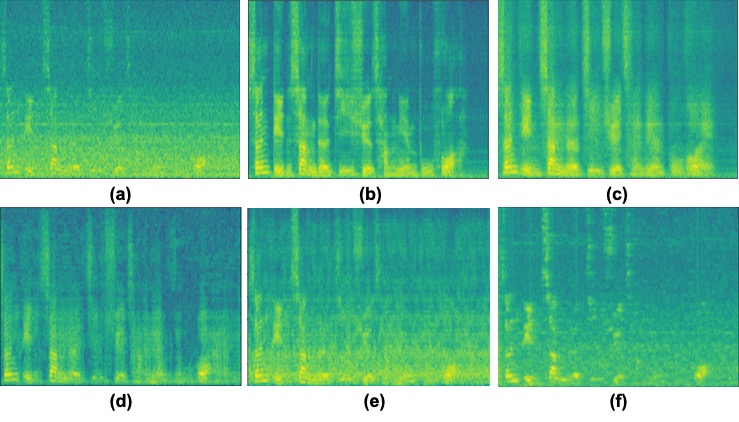}}
\caption{Spectrograms of the enhanced speech by different methods. (a) Noisy speech of engine noises at -1 dB. (b) Clean speech. (c)-(f) represent enhanced speech by our proposed method, LAVSE, AV-CVAE, and LogMMSE, respectively.}
\label{avse_result}
\end{figure}

\section{Experiments}
\label{sec:exp}

\subsection{Dataset and Training strategies}
This section presents the experimental setup and results. We evaluate our proposed models for the AVSE and AVSS tasks based on the TSMV dataset \footnote{\url {https://bio-asplab.citi.sinica.edu.tw/Opensource.html}}. The dataset contains video recordings of 18 native speakers (13 males and 5 females), each speaking 320 utterances
of Mandarin sentences. Each sentence consists of 10 Chinese characters, and the approximate duration for each utterance is 2–4 seconds. \\
\indent We optimized the AVSE and AVSS models with several training strategies. The first is partial fine-tuning (PF), which means that the weights of the feature extraction parts of AV-HuBERT are fixed, whereas the weights in the SSL block, that is, the transformer encoder, are updated based on the pre-trained checkpoint. Second, we investigate the entire fine-tuning (EF), whose difference from PF is the inclusion of weight updating for the feature extraction parts. Training from scratch (TFS) and training of the model without fine-tuning (WF) AV-HuBERT were also performed for comparison. The following sections detail the setup for the experiments for AVSE and AVSS, respectively.

\subsection{Audio-Visual SE Task}
\subsubsection{Experimental setup}
For the pre-processing parts in the videos, we cropped the mouth of the speakers via pre-trained CNN detectors~\cite{detect}. For the audio components, we followed the setup described in ~\cite{Chuang2022}. Of the 320 utterances for each speaker, we use the first 200 ones for training, and the remaining 120 ones are used for testing. To form clean-noisy speech pairs for training, the utterances were artificially corrupted by 100 types of noise~\cite{nonspeech} at five different signal-to-noise ratios (SNRs) ranging from \textminus{12} to 12 dB with an increment of 6 dB. The process generates approximately 600 hours of noisy utterances, and 12,000 noisy utterances are randomly selected to form a 9-hour training set to reduce the computation cost. To form the testing set, we selected five types of noise, including baby crying sounds, engine noise, pink noise, music noise, and street noise, with SNRs at \textminus{1} dB, \textminus{4} dB, \textminus{7} dB, and \textminus{10} dB.

\begin{figure}[t]
\centerline{\includegraphics[scale=.32]{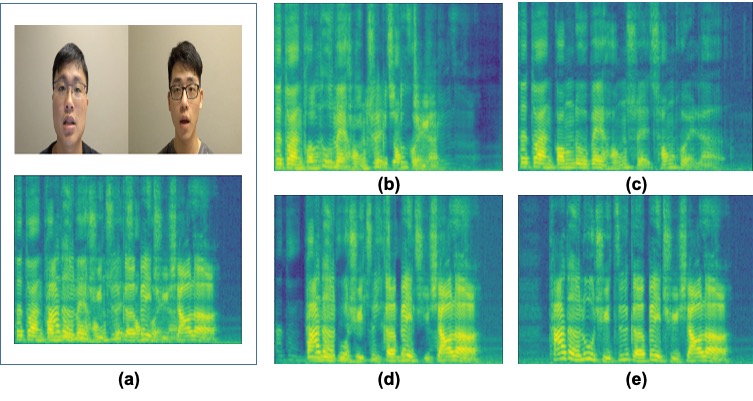}}
\caption{Results of our proposed AVSS framework. (a) A two-speaker talking video and the spectrogram of the mixed speech. (b) and (c) are spectrograms of the recovered speech for the left speaker and ground truth. (d) and (e) are spectrograms of the recovered speech for the right speaker and ground truth.}
\label{avss_result}
\end{figure}

\subsubsection{Model specification and training setup}
We used Base AV-HuBERT with 12 transformer layers pre-trained on the LRS3 dataset~\cite{Afouras18d} for five iterations as the checkpoint to build our AVSE system. The feature dimensions of the BLSTM model following AV-HuBERT were 256. For the fine-tuning process, the initial learning rate was set at 1e-4, and AdamW~\cite{loshchilov2018decoupled} was used as the learning optimizer. For each training strategy, we trained the model for 50 epochs and selected the model that yielded the best performance on the validation sets.

\subsubsection{Experimental results}
In this section, we compare the experimental results of the proposed AVSE model with those of other baseline SE models, including two AVSE models~\cite{Chuang2022,Sadeghi2020} and one traditional audio-only method, namely  LogMMSE~\cite{Loizou2005}. We conducted an objective comparison with two standardized evaluation metrics that are widely used to evaluate SE performance—the perceptual evaluation of speech quality (PESQ)~\cite{For01perceptualevaluation} and short-time objective intelligibility measure (STOI)~\cite{Taal_2011}. PESQ measures the quality of processed speech, whereas STOI is developed for evaluating speech intelligibility. PESQ and STOI scores range from \textminus{0.5} to 4.5 and 0 to 1, respectively. Higher scores indicate better performance.  \\
\indent Table~\ref{tab:avse} lists the PESQ and STOI scores for each SE method. From the table, we observe that the SE task is quite challenging because the traditional audio-only methods did not yield much improvement. In addition, the models using audio-visual information outperformed the audio-only methods, confirming the effectiveness of including visual formation in SE. Next, focusing on our proposed method, we can see that PF, EF, and WF all outperform TFS, indicating that the learned features in the pretrained AV-HuBERT model indeed result in a better AVSE system. We also implemented an AVSE system that used only the encoder part of the AV-HuBERT model (i.e., without the transformer part). These results are also reported in Table~\ref{tab:avse} and termed AVSE (w/o Trans.). From the table, we noticed that AVSE (w/o Trans.) does not perform as well as TFS does, suggesting that the transformer module is also effective for feature learning in the AVSE.\\
\indent Fig.\ref{avse_result} shows spectrograms of the enhanced speech from different SE approaches. We can observe the high-frequency components of the speech enhanced by partially fine-tuning the AV-HuBERT are better preserved than those from the LAVSE and AV-CVAE approaches, showing the consistency of the superiority of our approach. 

\begin{table}[h]
\centering
\begin{tabular}{lccc}
\hline
Methods & Mod. & PESQ & STOI \\
\hline
Noisy  &  & 1.18   & 0.60 \\
LogMMSE~\cite{Loizou2005} & (A)  & 1.21   & 0.61 \\
AV-CVAE~\cite{Sadeghi2020} & (AV)  & 1.34   & 0.63 \\
LAVSE~\cite{Chuang2022} & (AV)   & 1.31   & 0.61 \\
AVSE (w/o Trans.) & (AV) & 1.25  & 0.61 \\
AVSE (WF) & (AV)  & 1.30   & 0.63 \\
AVSE (TFS) & (AV)   & 1.26   & 0.60 \\
AVSE (EF) & (AV)  & 1.37   & 0.66 \\
\bf{AVSE (PF)} & (AV)  & \bf{1.40}   & \bf{0.68} \\
\hline
\end{tabular}
\caption{Results of speech enhancement via the audio-only or audio-visual methods.}
\label{tab:avse}
\end{table}

\subsection{Audio-Visual Speech Separation}
\subsubsection{Experimental setup}
AVSS experiments were based on the TMSV dataset. As shown in Fig.\ref{avss_result}(a), we made a two-talker video by merging two individual videos side by side, while redoing the soundtrack by mixing the original ones. The spectrogram depicted in Fig.\ref{avss_result}(a) is the mixed speech from the utterances of the left speaker, as shown in Fig.\ref{avss_result}(c), and the right speaker, as shown in Fig.\ref{avss_result}(e). We randomly selected two utterances from two different speakers to form a mixed speech for separation. A total of 12,000, 1,200, 1,200 mixed utterances for training, validation and testing were made with spoken sentences and speakers mismatched.

\subsubsection{Model specification and training setup}
We used the same AV-HuBERT checkpoint as that used for the AVSE task. For the multi-head attention modules, we used 12 heads with a feature dimension of 512. The neural regression module of the AVSS model was the same as that used in the AVSE model. For the fine-tuning process, the initial learning rate, optimizer, and training epochs were the same as those for the AVSE task.

\subsubsection{Experimental results}
Fig.\ref{avss_result} shows spectrograms of the mixed speech and the separated speech by the proposed AVSS approach. Fig.\ref{avss_result}(a) demonstrates a two-speaker sample video where the speakers are speaking simultaneously. The spectrogram of the mixed speech in Fig.\ref{avss_result}(a) is composed of the clean utterances from the left speaker and the right speaker. The spectrograms of the separated speech obtained by fine-tuning AV-HuBERT of our AVSS model are presented in Figs.\ref{avss_result} (b) and (d), with the respective ground-truth Figs.\ref{avss_result} (c) and (e). We can observe from the spectrograms that our method can effectively separate mixed speech using the corresponding visual information.  \\
\indent To quantitatively assess the results of different training strategies, we used two metrics for evaluation: the scale-invariant signal-to-noise ratio (SI-SNR) and source-to-distortion ratio (SDR). These two metrics are common for evaluating separated speech~\cite{Luo2018}. Table~\ref{tab:avss} reports the two scores of the separated speech using the different training strategies of our AVSS methods. The scores related to PF were the best among the training strategies, suggesting that the learned representations of AV-HuBERT can be effective for the AVSS task as well. Nevertheless, note that in the AVSS task, TFS is only second to PF, unlike AVSE, where the performance of TFS is worse than that of PF, EF, and WF, as reported in Table.\ref{tab:avse}. We argue that including multiple attention heads improves the coupling of multimodal features from different speakers, thereby closing the performance gap between TFS and the others via leveraging the pre-trained AV-HuBERT features.

\begin{table}[h] 
\centering
\begin{tabular}{lcc}
\hline
Methods & SISNR (dB) & SDR (dB) \\
\hline
AVSS (WF)   & 3.03   & 4.08 \\
AVSS (TFS)   & 3.53   & \bf{4.63} \\
AVSS (EF)   & 3.27   & 4.31 \\
AVSS  (PF)   & \bf{3.59}   & 4.59 \\
\hline
\end{tabular}
\caption{Results of audio-visual speech separation by different fine-tuning strategies.}
\label{tab:avss}
\end{table}

\section{Conclusion}
\label{sec:conclusion}
In this study, we proposed novel AVSE and AVSS frameworks that leveraged a pretrained AV-HuBERT model. For AVSE, we demonstrated that after partially fine-tuning the AV-HuBERT model, our AVSE system outperformed other baseline SE models. For AVSS, we noted similar trends and the advantages of using AV-HuBERT embeddings. In summary, this study demonstrated how a pre-trained AV-HuBERT model can improve the training of AVSE and AVSS tasks, showing its promising ability for AV regression tasks.

\bibliographystyle{IEEEbib}
\bibliography{strings,refs}

\end{document}